\documentclass[prl,twocolumn,aps,nofootinbib,nobibnotes,superscriptaddress,
preprintnumbers]{revtex4-1}
\usepackage{epsfig}
\usepackage{graphics}
\usepackage{bm}
\usepackage{color}

\newcommand\ee{{\boldsymbol \varepsilon}}

\def\ba{\begin{eqnarray}}
\def\ea{\end{eqnarray}}
\def\bea{\begin{eqnarray}}
\def\eea{\end{eqnarray}}
\def\be{\begin{equation}}
\def\ee{\end{equation}}
\def\d{\mathrm{d}}
\def\({\left(}
\def\){\right)}
\def\[{\left[}
\def\]{\right]}

\begin{document}

\title{Cosmic Expansion in Extended Quasidilaton Massive Gravity}

\date{\today,~ $ $}
\preprint{NORDITA-2014-139}

\author{Tina Kahniashvili}
\email{tinatin@andrew.cmu.edu}
\affiliation{The McWilliams Center for Cosmology and Department of Physics, Carnegie Mellon University, 5000 Forbes Ave, Pittsburgh, PA 15213, USA}
\affiliation{Department of Physics, Laurentian University, Ramsey Lake Road, Sudbury, ON P3E 2C, Canada}
\affiliation{Abastumani Astrophysical Observatory, Ilia State University, 3-5 Cholokashvili Ave, Tbilisi, GE-0194, Georgia}

\author{Arjun Kar}
\email{arjunkar@andrew.cmu.edu}
\affiliation{The McWilliams Center for Cosmology and Department of Physics, Carnegie Mellon University, 5000 Forbes Ave, Pittsburgh, PA 15213, USA}

\author{George Lavrelashvili}
\email{lavrela@itp.unibe.ch }
\affiliation{Department of Theoretical Physics, A.Razmadze Mathematical Institute \\ I.Javakhishvili Tbilisi State University, GE-0177 Tbilisi, Georgia}
\affiliation{Max-Planck-Institute for Gravitational Physics \\ Albert-Einstein-Institute, D-14476 Potsdam, Germany}

\author{Nishant~Agarwal}
\email{nua11@psu.edu}
\affiliation{Department of Astronomy and Astrophysics, The Pennsylvania State University, University Park, PA 16802, USA}
\affiliation{Institute for Gravitation and the Cosmos, The Pennsylvania State University, University Park, PA 16802, USA}

\author{Lavinia Heisenberg}
\email{laviniah@kth.se}
\affiliation{Nordita, KTH Royal Institute of Technology and Stockholm University, Roslagstullsbacken 23, 10691 Stockholm, Sweden}
\affiliation{Department of Physics \& The Oskar Klein Centre, AlbaNova University Centre, 10691 Stockholm, Sweden}

\author{Arthur Kosowsky}
\email{kosowsky@pitt.edu}
\affiliation{Department of Physics and Astronomy, University of Pittsburgh, 3941 O'Hara Street, Pittsburgh, PA 15260 USA}
\affiliation{Pittsburgh Particle Physics, Astrophysics, and Cosmology Center (PITT PACC), Pittsburgh, PA 15260 USA}

\date{\today}

\begin{abstract}

Quasidilaton massive gravity offers a physically well-defined gravitational theory with non-zero graviton mass. We present the full set of dynamical equations governing the expansion history of the universe, valid during radiation domination, matter domination, and a late-time self-accelerating epoch related to the graviton mass. The existence of self-consistent solutions constrains the amplitude of the quasi-dilaton field and the graviton mass, as well as other model parameters. We point out that the effective mass of gravitational waves can be significantly larger than the graviton mass, opening the possibility that a single theory can explain both the late-time acceleration of cosmic expansion and modifications of structure growth leading to the suppression of large-angle correlations observed in the cosmic microwave background.

\end{abstract}

\maketitle


\noindent
{\bf Introduction.}
In the standard cosmological model, the acceleration of the universe is assumed to be due to a cosmological constant $\Lambda$ that has time-independent energy density which becomes dominant at late times. An alternative possibility is that the true theory of gravity differs from general relativity on large scales.

One interesting physical modification of general relativity is a non-zero graviton mass $m_g$. However, constructing a consistent non-linear theory of massive gravity has been a challenging task. At the linearized level, the action must have the Fierz-Pauli form \cite{Fierz:1939ix} to be ghost-free, but any purely linear theory suffers from the van Dam-Veltman-Zakharov (vDVZ) discontinuity \cite{vanDam:1970vg,Zakharov:1970cc}: the theory does not reduce to general relativity in the limit of zero graviton mass. Non-linear extensions to the Fierz-Pauli theory can incorporate a strong-coupling phenomenon known as the Vainshtein mechanism \cite{Vainshtein:1972sx} which evades the vDVZ issue in the vicinity of matter sources. But these theories were soon found to contain an unhealthy ghost-like degree of freedom \cite{Boulware:1973my,Dubovsky:2004sg}.

Recently, theories of ghost-free massive gravity (the dRGT theory) \cite{deRham:2010ik,deRham:2010kj} and its bigravity generalization \cite{Hassan:2011zd} have been discovered. Pioneering work on cosmological aspects of these theories  showed that the original dRGT theory does not allow isotropic and homogeneous background solutions \cite{D'Amico:2011jj}, but that massive gravity could provide a cosmological constant-like term in the field equations leading to accelerated cosmological expansion \cite{Gratia:2012wt}. Searches for a fully satisfactory theory have led to models of massive gravity incorporating a quasidilaton field \cite{D'Amico:2012zv,DeFelice:2013tsa,DeFelice:2013dua,Mukohyama:2014rca}; the extra field is used to stabilize the metric perturbations. Cosmological perturbations in extended theories of massive gravity have been studied, for example, in Refs.~\cite{Gumrukcuoglu:2013nza, Gumrukcuoglu:2014xba, Mukohyama:2014rca}. For recent reviews see Refs.~\cite{deRham:2014zqa,Tolley:2015oxa}.

Here we present the full dynamical background equations governing the expansion history of the universe in quasidilaton massive gravity, valid at all epochs. Recent work has assumed the late-time asymptotic form of the expansion history to evaluate the growth of perturbations \cite{Motohashi:2014una} but the full solution is required for comparison with observations, since our current epoch is in the transition between matter domination and accelerating expansion. We find constraints on parameters in the model arising from the requirement of self-consistent cosmological solutions with late-time acceleration. We also note that the mass scale governing the evolution of tensor perturbations can be substantially larger than $m_g$; if a similar scale governs the evolution of scalar perturbations as well, then the observed lack of correlations at large angular scales in the microwave background \cite{Copi:2010na} might also arise in these theories.

\


\noindent
{\bf Extended Quasidilaton Massive Cosmologies.}
Here we analyze the extended quasidilaton theory of Ref.~\cite{DeFelice:2013tsa}, defined by the action
\begin{equation}
	S = S_Q + S_M \, ,
\end{equation}
where
\begin{eqnarray} \label{action}
	S_Q & = & \frac{M_{\rm Pl}^2}{2}\int {\rm d}^4x\sqrt{-g} \bigg[ R - 2\Lambda - \frac{\omega}{M_{\rm Pl}^2} \partial_\mu \sigma \partial^\mu \sigma \nonumber \\
	& & \quad + \ 2m^2_g (\mathcal{L}_2 + \alpha_3\mathcal{L}_3 + \alpha_4\mathcal{L}_4 ) \bigg] \, ,
\end{eqnarray}
and $S_M$ is the matter action, which we will assume to be that for a perfect fluid. Here $M_{\rm Pl}$ is the Planck mass, $\Lambda$ is a cosmological constant, $\sigma$ (taken to be $=\bar{\sigma}(t)$) is the quasidilaton field, and ${\mathcal L}_2 - {\mathcal L}_4$ are the terms that provide a mass $m_g$ to the graviton. The theory depends on $m_g$ and additionally four dimensionless coupling constants $\omega, \, \alpha_{\sigma}, \, \alpha_3$, and $\alpha_4$. The physical and fiducial metrics are characterized through the following functions in FLRW space-time: the Hubble parameter $H(t) \equiv {\dot a}/a$, $a(t)$ being the scale factor, and a dot representing the derivative with respect to physical time $t$,  the quasidilaton field ${\bar\sigma}(t)$ and a convenient related variable $X(t) \equiv \exp(\bar\sigma/M_{\rm Pl})/a$; the St\"{u}ckelberg fields $\phi^a$ (we work with $\phi^0 = \phi^0(t)$ and $\phi^i = x^i$); and $q(t) \equiv n(t) a(t)$, $n(t)$ being the lapse function of the extended fiducial metric, expressed in terms of $\dot{\phi}^0$ and ${\bar\sigma}(t)$ by
\begin{equation}
	\left(\frac{\dot{\phi}^0}{n}\right)^2 \equiv 1 - {\alpha_\sigma}\left(\frac{\dot{\sigma}}{q X m_g}\right)^2 \, ,
\label{nequation}
\end{equation}
where we have redefined $\sigma$ to denote ${\bar \sigma}/M_{\rm Pl}$.

Equations of motion for the expansion history $a(t)$ and field values can be obtained by varying the action, $S$. Introducing the auxiliary function $J\equiv3+3\alpha_3 (1-X) +\alpha_4 (1-X)^2$, variation with respect to $\phi^0$ gives the constraint equation
\begin{equation} \label{eq:phi_0}
	\frac {{\rm d}}{{\rm d}t} \left[a^4 X (1-X) J \frac{\dot{\phi}^0}{n} \right] =0 \, ,
\label{constraint}
\end{equation}
while variations with respect to $N(t)$\footnote{Note that we work in the proper time gauge, in which the lapse function $N(t)$ of the physical metric is
$N = 1$. The variation, of course, has to be done before gauge fixing.} and $a(t)$ give the modified first and second Friedmann equations,
\begin{eqnarray}
	3 H^2 - \frac{\omega}{2} {\dot \sigma}^2 = \Lambda + \Lambda_X +\frac{1}{M_{\rm Pl}^2} \rho_M \, ,
\label{friedman1} \\
	\dot{H} = \frac{1}{6} (q-1) \frac{{\dot \Lambda}_X}{{\dot \sigma} -H} - \frac{{\omega}}{2}{\dot \sigma}^2 - \frac{1}{2 M_{\rm Pl}^2} (\rho_M+P_M) \, .
\label{friedman2}
\end{eqnarray}
Here $\Lambda_X = m_g^2 (X-1) [J +(X-1)(\alpha_3(X-1) -3)]$ is the effective cosmological constant, ${\dot\Lambda}_X = 3m_g^2 {\dot X} \left[ J + X(\alpha_3 (X-1) -2)\right]$ and $\dot\sigma = H +\dot X/X$. The total energy density $\rho_M$ and pressure $P_M$ for the matter terms satisfy the usual continuity equation $\dot{\rho}_M+3 H (\rho_M+P_M) = 0$. The functions $H$ and $X$ are connected through the dynamical relation\footnote{We thank Emir G\"{u}mr\"{u}k\c{c}\"{u}o\u{g}lu for pointing out an error in eq.\ (\ref{y}) of the previous version of our paper. The corresponding quasidilaton equation of motion now matches that given in \cite{Gumrukcuoglu:2016hic}.}
\begin{widetext}
\bea
	m_g^2 q \frac{X}{H+\dot{X}/X} \[ 4(X-1)JH + \frac{\dot{X}}{X} \{ 3X\( \alpha_3(X-1)-2 \) + J(4X-1) \}\] - \frac{\alpha_{\sigma}}{a^4X} \frac{\d}{\d t} \[ \frac{a^4}{q} \(H+\frac{\dot{X}}{X}\) (X-1) J \] & = & 0 \, , \nonumber \\
\label{y}
\eea
\end{widetext}
which imposes constraints on the allowed model parameters. In the following we will restrict our analysis to $\dot{\phi}^0(t)/n(t) = {\rm constant}$, which follows from Eqs.~(\ref{constraint}) and (\ref{y}), if we ignore the $\alpha_{\sigma}$ term in Eq.~(\ref{y}). Then using Eq.~(\ref{nequation}) we obtain ${\dot q}/q = {\ddot \sigma}/{\dot \sigma} - {\dot X}/X $ if $H \neq - \dot X/X$. Finally, the inequality $({\dot{\phi}^0}/{n})^2 > 0$ implies that $0 < \alpha_\sigma (H+\dot X/X)^2 < q^2 X^2 m_g^2$.

\


\noindent
{\bf Self-Accelerating Regime.}
A late-time attractor solution for this system exists \cite{D'Amico:2012zv,DeFelice:2013tsa}, and is characterized by a constant Hubble parameter $H_A\equiv H(a\rightarrow\infty)$, and constant values of $X_A \equiv X(a \rightarrow \infty)$ and $q_A \equiv q(a\rightarrow \infty)$. In this limit $\rho_M$ and $P_M$ vanish. We will consider solutions for which $J \rightarrow 0$ as $a \rightarrow \infty$. The parameter $\alpha_{\sigma}$ does not affect the background dynamics, but plays a crucial role in constructing a healthy theory: linear stability in the asymptotic regime implies that \cite{DeFelice:2013tsa} $1 < \alpha_{\sigma} [H_A/(X_A m_g)]^2< q_A^2 $, which can be achieved by a proper choice of $\alpha_{\sigma}$ if $q_A > 1$.

The asymptotic limits of Eqs.~(\ref{friedman1}) and (\ref{friedman2}) relate $H_A$, $X_A$, and $q_A$ --- setting the background cosmological constant $\Lambda$ to zero we obtain \cite{DeFelice:2013tsa}
\begin{eqnarray}
	\left(3 -\frac{\omega}{2} \right) H^2_A = \Lambda_A \, ,
\label{HA} \\
	q_A-1 = \frac{\omega H^2_A }{m_g^2 X_A^2 [\alpha_3 (X_A-1) -2]} \, ,
\label{rA}
\end{eqnarray}
where $\Lambda_A \equiv \Lambda_X (a \rightarrow \infty) = m_g^2 (X_A-1)^2 [\alpha_3(X_A-1) -3]$. The requirement that the effective Newton's constant should be positive gives the constraint $\omega < 6$ \cite{DeFelice:2013tsa}. Note that Eqs. (\ref{HA}) and (\ref{rA}) are valid only in the limit $a \rightarrow \infty$ and cannot be used as dynamical equations or to describe the coupling with matter, when $\Lambda_X \leq \rho_M/M_{\rm Pl}^2$: the non-zero functions $J(a)$ and ${\dot X}(a)$ must be included in the dynamics until the asymptotic limit ($a \rightarrow \infty$)  is reached.

The condition $J=0$ is simply an algebraic equation for $X$ with the roots $2 \alpha_4 X_{A \pm} = {3\alpha_3 + 2 \alpha_4 \pm \sqrt{9{\alpha_3}^2 -12\alpha_4}}$. Requiring that $X_A$ is real gives $\alpha_4 < \frac{3}{4} \alpha_3^2$, while $\Lambda_X > 0$ implies that $\alpha_3(X_A-1) >3 $.
Values $X_A<1$ are realized only for negative values of the parameter $\alpha_3$. Eqs. (\ref{HA}) and (\ref{rA}) give the asymptotic relation
\begin{equation}
	q_A= 1 + \frac{\omega H_A^2}{m_g^2 X_A^2 \left(1+ \frac{\left( 3- \frac{\omega}{2} \right) H_A^2}{m_g^2 (X_A-1)^2}\right)} \, .
\end{equation}
Enforcing the condition $\omega < 6$ gives ${m_g^2 X_A^2}/{H_A^2} < {6}/(q_A-1)$.

We also have an additional constraint coming from tensor perturbations, which are characterized by the effective mass of gravitational waves, $M_{\rm GW}(a)$. For stability, we require that $M_{\rm GW}^2 > 0$. In the asymptotic limit $a \rightarrow \infty$, $M_{\rm GW}^2$ tends towards a constant value (see Eq. (38) of Ref. \cite{DeFelice:2013tsa}), expressed in terms of model parameters $\omega$ and $m_g$, and dynamically generated parameters $X_A$ and $H_A$ as
\begin{eqnarray}
	& & \frac{M_{\rm GW}^2}{H_A^2}(a\rightarrow \infty) = X_A^2 \left[ \frac{m_g^2}{H_A^2} +\frac{3-\frac{\omega}{2}}{(X_A-1)^2} \right] \nonumber \\
	& & \quad \quad  + \frac{m_g^2}{H_A^2} \frac{\omega X_A (X_A-1)}{\left(3-\frac{\omega}{2}\right) + \frac{m_g^2}{H_A^2} (X_A-1)^2}
+ \omega \frac{X_A+1}{X_A-1} \, . \quad \quad
\label{M3}
\end{eqnarray}
For $0 < X_A < 1$  the condition $M_{\rm GW}^2 >0$ implies that $1 < q_A < \bar{q}_A$, with
\begin{eqnarray}
	\bar{q}_A & = & \frac{2}{1+X_A}  \nonumber \\
	& - & \frac{\omega}{X_A (1+X_A)}
\left[\frac{\frac{m_g}{H_A} (X_A-1)^2}{\left(3 - \frac{\omega}{2}\right) + \frac{m_g^2}{H_A^2} (X_A-1)^2}\right]^2 \, . \quad \ \
\label{barq}
\end{eqnarray}
To summarize, assuming that $\alpha_{\sigma}$ is chosen to satisfy the linear stability condition, the first stability island is characterized by $0 < X_A < 1$, $1 < q_A < \bar{q}$, $ 0 < \omega < 6$. For $X_A>1$ the mass function is always positive (assuming that $q_A>1$ is satisfied) and we do not get any additional constraints.

\begin{figure}[t!]
\includegraphics[width=0.87\columnwidth]{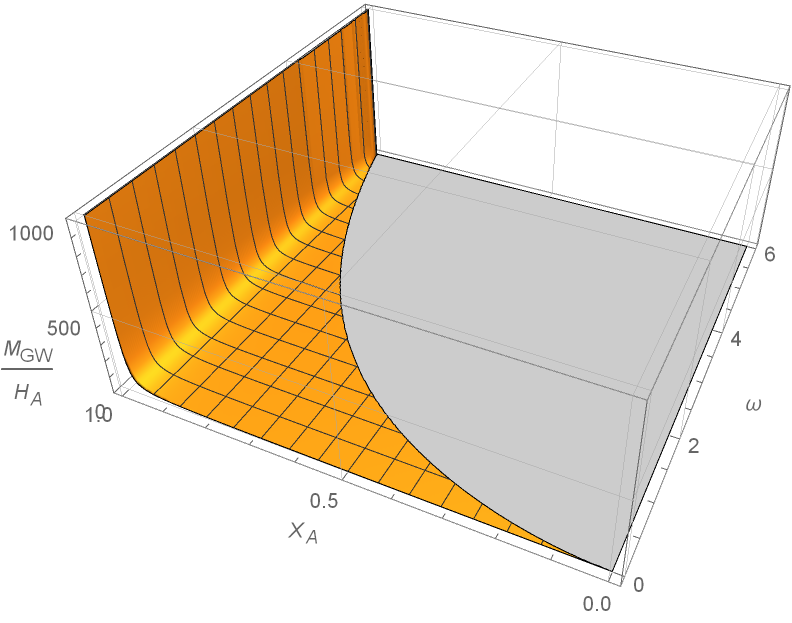}
\includegraphics[width=0.87\columnwidth]{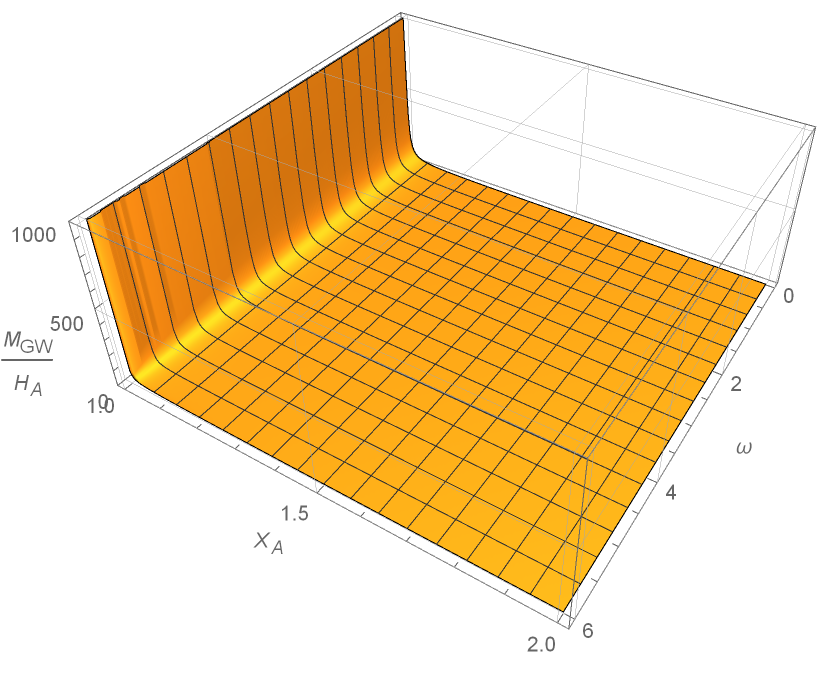}
\caption[]{(Color online) Plot of $M_{\rm GW}/H_A$ for $m_g = H_A$ in the entire region for $0<X_A<1$ (upper panel) and for $X_A>1$ (lower panel). The excluded regions (which correspond to negative $M^2_{\rm GW}/H^2_A$) are shown in grey color.
}
\label{parameters1}
\end{figure}

The allowed parameter ranges are presented in Fig. \ref{parameters1}. In particular, we see that there are parameter ranges for which $M_{\rm GW}/H_A \gg 1$ even if $m_g/H_A \simeq 1$.
This can understood analytically by observing that in the limit of small $|X_A-1|$, $M^2_{\rm GW}/H^2_A \approx (3-\omega/2)/(X_A-1)^2 + {\cal O}[X_A-1]^{-1}$. Adjusting the value of the parameter $\alpha_3$, we can have, at the same time, a sizeable value of the effective cosmological constant $\Lambda_A$. This is an important property of quasidilaton extensions of massive gravity. In massive gravity, the constraint $m_g/H_A \simeq 1$ severely restricts $M_{\rm GW}$, however, in the presence of the quasidilaton this restriction is relaxed and one can have $M_{\rm GW}/H_A \gg 1$, with important phenomenological implications.

\


\noindent
{\bf Radiation and Matter Dominated Regimes.}
In order to recover the usual background expansion during matter (or radiation) domination, the effective cosmological constant $\Lambda_M$ as well as the quasidilaton field kinetic term $\omega \dot\sigma_M^2/2 $ must be negligible compared to the total energy density of matter: $\Lambda_M \ll \rho_M/M_{\rm Pl}^2 \simeq 3H_M^2$ and $\omega \dot\sigma_M^2/2 \ll 3H_M^2 \simeq \rho_M/M_{\rm Pl}^2$ (the subscript $M$ denotes matter or radiation domination). Using Eqs. (\ref{friedman1}) - (\ref{friedman2}) we obtain the following (dynamical) equation for $q_M(t)$,
\begin{eqnarray}
	& & q_M \ = \ 1 + \frac{\omega H_M^2}{m_g^2 X_M}  \nonumber \\
	& & \ \times \, \frac{ [J_M+X_M(\alpha_3 (X_M-1) -2)] }{\big([J_M+X_M(\alpha_3 (X_M-1) -2)] +\frac{4}{3} J_M (X_M-1)\big)^2} \, . \nonumber \\
\label{rM}
\end{eqnarray}
This equation can be used when formulating the initial conditions to determine the background expansion history.
Note that $3 \omega {\dot X}_M {\dot\sigma}_M^2 = (q_M-1) {\dot\Lambda}_M X_M$ (see Eq. (\ref{friedman2}) with the use of  $2 {\dot H}_M = -(\rho_M+P_M)/M_{\rm Pl}^2$). Eq.~(\ref{constraint}) with ${\dot\phi}^0(t)/n(t) = $ constant results in the following scaling $J_M X_M (1-X_M) \propto 1/a^4$.

\


\noindent
{\bf Imprints of Massive Gravitational Waves on the Microwave Background.}
In the linear and Lorentz-invariant massive gravity theory \cite{Fierz:1939ix,Eardley:1973br}, gravitons are spin-2 particles and there are five independent polarizations of gravitational waves, usually referred to as helicity states $\pm 2$, $\pm 1$, and $0$. The additional degrees of freedom (helicities $\pm 1$ and $0$) can be associated with vector and scalar modes (i.e.\ vector and scalar particles). The vDVZ discontinuity arises because of the coupling of the spin-0 mode with matter, and it can be eliminated, as stated before, through the Vainshtein screening mechanism. Another difficulty of many massive gravity models is the presence of ghosts in the models. As noted earlier, the recently proposed dRGT model \cite{deRham:2010kj} is a ghost-free model of massive gravity. In this framework one can define the cross-over density $\rho_{\rm cr} = 3M_{\rm Pl}^2 m_g^2$; when the density of the universe is lower than $\rho_{\rm cr}$, the effects of a non-zero graviton mass start to become relevant.

We consider here only tensor modes (gravitational waves) which acquire a non-zero mass $M_{\rm GW}$, and the gravitational wave dispersion relation is modified accordingly, $f^2 = (k/a)^2 +M_{\rm GW}^2$, $k$ being the comoving wavenumber and $f$ the effective frequency with respect to proper time. As in GR the tensor mode of perturbations is generated through quantum-mechanical fluctuations during inflation \cite{Grishchuk:1974ny}. In the standard cosmological scenario the tensor mode is amplified through the mechanism of parametric resonance \cite{Starobinsky:1979ty,Rubakov:1982df}, and the amplitude is completely determined by the value of the Hubble parameter at the end of inflation. The content of the universe as well as the current accelerated cosmological expansion slightly affect the predicted background of gravitational waves in the present epoch, even in GR \cite{Zhang:2005nw}. In massive gravity models the situation is more complicated. The oscillator equation for gravitational waves contains an additional mass term $a^2 M_{\rm GW}^2$, which affects the lower frequencies $(k/a) \lesssim M_{\rm GW}$, leaving the higher frequency tail of gravitational waves almost unchanged. The largest deviations from GR are therefore expected for these low frequency modes, and the authors in Ref. \cite{Gumrukcuoglu:2012wt} in fact found a sharp peak in the gravitational wave spectrum around $f_0 \simeq M_{{\rm GW},0}$, the subscript 0 denoting the values today.

We would like to conclude this section with some speculations for the microwave background {\it scalar} power spectrum. For an effective mass scale $m$ associated with gravity waves, we might expect a Yukawa-like gravitational potential of the form, $V(r) \propto e^{-mr}/r$. A natural choice for the mass scale $m$ in a massive gravity theory is probably between $m_g$ and $M_{\rm GW}$. If we take $m$ to be as large as $M_{\rm GW}$ then we expect to see a suppression in power on scales similar to $M_{\rm GW}^{-1}$. In this paper we pointed out that in a quasidilaton theory of massive gravity, $M_{\rm GW}$ could be much larger than the Hubble scale. So we can expect two scales to appear in the data; if we choose, for example, $m_g \simeq H_0$ and $M_{\rm GW} \simeq 1000H_0$, then we can relate the first scale to the accelerated expansion of the universe and the second one to the large-scale suppression of power in the microwave background two-point correlation function \cite{Copi:2010na}. Note that this freedom of having two distinct scales might not be available in ``pure'' massive gravity models (with no additional fields) since there we do not expect $M_{\rm GW}$ to be very different from $m_g$ \cite{D'Amico:2011jj,Gumrukcuoglu:2013nza}.

\


\noindent
{\bf Concluding Remarks.}
Massive gravity and its quasidilaton extension provide a very interesting theoretical framework and have a rich phenomenology. In this paper we investigated some cosmological consequences of quasidilaton massive gravity. We presented the general background equations which are valid in both asymptotic regimes --- matter domination and the current self-accelerated stage. These dynamical equations are needed in order to compare with the full expansion history of the universe.

We also showed that the quasidilaton model permits solutions with $m_g/H_A \simeq 1$ and $M_{\rm GW}/H_A \gg 1$. This allowed us to speculate that the quasidilaton model may provide a way to explain the large-angle suppression of power in the microwave background, in addition to the current accelerated expansion of the universe. In order to study this further one needs to calculate the detailed perturbation equations, which can then be compared directly with cosmological data on the microwave background. We plan to report on this in the future.

\


\noindent
{\bf Acknowledgments.}
We appreciate useful discussions with Raphael Flauger, Gregory Gabadadze, Emir G\"{u}mr\"{u}k\c{c}\"{u}o\u{g}lu, Sayed Hassan, and Tomi Koivisto. We acknowledge partial support from the Swiss NSF SCOPES Grant IZ7370-152581, the CMU Berkman foundation, the NSF Grants AST-1109180 and AST-1312380, and the NASA Astrophysics Theory Program Grant NNXlOAC85G.


\end{document}